\documentclass[12pt, preprint]{aastex}

\begin{document}
 
\title{Gravitational Lensing by Burkert Halos}
 
\author{Yousin Park\altaffilmark{1,2} and Henry C. Ferguson\altaffilmark{2,1}}
\email{ypark@stsci.edu, ferguson@stsci.edu}
 
\altaffiltext{1}{Department of Physics and Astronomy, The Johns Hopkins University, Baltimore, MD
21218}
\altaffiltext{2}{Space Telescope Science Institute, Baltimore, MD 21218}
 
\begin{abstract}
 
We investigate the gravitational lensing properties of dark matter
halos with Burkert profiles. We derive an analytic expression for the
lens equation and use it to compute the magnification, impact parameter
and image separations for strong lensing. For the scaling relation that
provides the best fits to spiral-galaxy rotation curve data, Burkert
halos will not produce strong lensing, even if this scaling relation
extends up to masses of galaxy clusters. 
Tests of a simple model of an exponential stellar disk
superimposed on a Burkert-profile halo demonstrate that strong lensing
is unlikely without an additional concentration of mass in the galaxy
center (e.g. a bulge). The fact that most strong lenses on galactic
scales are elliptical galaxies suggests that a strong central concentration
of baryons is required to produce image splitting.
This solution is less attractive for clusters of galaxies, which are
generally considered to be dark-matter dominated even at small radii. 
There are three possible implications of these results: 
(1) dark halos may have a 
variety of inner profiles (2) dark matter halos may not follow a 
single scaling relation from galaxy scale up to cluster scale and/or
(3) the splitting of images (even by clusters of galaxies) may in general be 
due to the central concentration of baryonic material in halos 
rather than dark matter.

\end{abstract}

%\keywords{gravitational lensing---galaxies: halos---cosmology: dark matter}

\section{Introduction}

It is currently believed that structure in the universe formed with
nearly scale-free initial conditions from quantum fluctuations during
the early parts of the inflationary phase. Subsequent growth of
structure depends in part on the nature of the dark matter (DM);
specifically if it is cold (non-relativistic at the epoch of
matter--radiation decoupling), as would be the case for axions,
clustering proceeds hierarchically with small structures collapsing
before large ones.  The exact nature of this evolution depends on the
cosmological parameters and the nature of the DM. If the DM is entirely collisionless and cold, halos are expected to form with a nearly universal profile, independent of the final mass of the
halo. A specific functional form for this profile was proposed by
Navarro, Frenk \& White (1996, hereafter NFW) on the basis of 
N-body simulations of hierarchical structure formation.

On a galaxy/sub-galaxy scale, the NFW profile for DM halos has been
severely challenged by observations of rotation curves of disk galaxies
(e.g.  de Blok et al. 2001; Borriello \& Salucci 2001). The rotation
curves of low-surface-brightness (LSB), low-mass galaxies tend to rise
significantly more slowly with radius than the NFW prediction. The
Burkert profile for DM halos was proposed to explain the rotation
curves of four dwarf galaxies (Burkert 1995), and appears to provide an
excellent mass model for DM halos around disk systems up to 100 times
more massive (Salucci \& Burkert 2000). For the DM halos haboring
elliptical galaxies, Keeton (2001) shows that the NFW profile is too
concentrated to explain observed gravitational lensing. Recently,
Borriello, Salucci \& Danese (2002) show that the NFW profile is
inconsistent with the Fundamental Plane (FP), and that the FP can be
successfully explained with the Burkert profile. On a cluster scale, Wu
\& Xue (2000) find that the Burkert and NFW profiles provide almost
indistinguishable fits to ROSAT cluster X-ray data. All this evidence
suggests that the Burkert profile is a $plausible$ candidate for the DM
halos on both galaxy and cluster scales, though it was empirically
proposed and its physical origin is less well understood than
the NFW profile.

Gravitational lensing offers a promising way to distinguish the
different proposed halo profiles.  In this letter we present an analytical
expression for the Burkert-profile lens equation and show that
this profile will produce strong lensing only for a certain
combination of concentration index and lens--source distances.  
We find that Burkert halos of any mass are unlikely to act as 
strong lenses unless the spiral-galaxy scaling relation between characteristic
mass and density breaks down on larger mass scales, or other
(e.g. baryonic) components of the object provide the cusp necessary 
for image splitting.

In $\S$ 2, we present the lens equation for the Burkert
profile. In $\S$ 3, we explore the properties of the Burkert lens and
examine the implications of these properties. In $\S$ 4, we summarize and discuss the results. In
this letter, we consider $\Lambda$CDM model:  $\Omega_M$ = 0.3 and
$\Omega_{\Lambda} = 0.7$ with a Hubble constant $H_0 = 100 h {\rm
km/s/Mpc}$ and $h = 0.75$.

\section{Lens Equation for the Burkert Profile}

The Burkert profile has a constant density core (Burkert 1995):
\begin{equation}
\rho(r) = \frac{\rho_s}{(1 + r / r_s) [1 + (r / r_s)^2]},
\end{equation}
where $\rho_s$ and $r_s$ are free parameters that represent the central
DM density and the scale radius. The NFW profile has
singularity at its center:
\begin{equation}
\rho_{{\rm NFW}}(r) = \frac{\rho_s}{r / r_s (1 + r / r_s)^2}.
\end{equation}
At very large radii, two profiles coincide: $\rho \sim r^{-3}$.

In the thin lens approximation, the 
lens equation for an axially symmetric mass profile is essentially one dimensional:

\begin{equation}
\psi({\theta}) = \theta - \frac{D_{LS}}{D_S} \frac{4GM_p(D_L|\theta|)}{c^2 D_L \theta},
\end{equation}
where $D_{LS} = D_A(z_L, z_S)$, $D_L = D_A(0, z_L)$ and $D_S = D_A(0,
z_S)$. Here, $z_L$ and $z_S$ are respectively the redshifts of the lens
and the source, and $D_A(z_1, z_2)$ is the angular-diameter distance
from $z_1$ to $z_2$. And, $\psi$  and $\theta$ are respectively the
positions of the source ($\psi \ge 0$) and the image, and $M_p(\xi)$ is
the projected mass enclosed within a projected radius $\xi$:

\begin{equation}
M_p(\xi) = 2 \pi \int_0^{\xi} x dx \int_{-\infty}^{\infty} dz \rho(\sqrt{x^2 + z^2}) = 4 \pi \rho_s r_s^3 a(\frac{\xi}{r_s}).
\end{equation}
The function $a(x)$ can be worked out analytically:
\begin{equation}
a(x) = \left \{ \begin{array}{lc}
\ln \frac{x}{2} + \frac{\pi}{4}(\sqrt{x^2 +1} - 1) + \frac{\sqrt{x^2 + 1}}{2} {\rm arccoth} \sqrt{x^2 + 1} - \onehalf \sqrt{x^2 - 1} \arctan \sqrt{x^2 - 1} & (x > 1) \\ -\ln 2 - \frac{\pi}{4}  + \frac{1}{2 \sqrt{2}} [\pi + \ln (3 + 2 \sqrt{2})] & (x = 1) \\ \ln \frac{x}{2} + \frac{\pi}{4}(\sqrt{x^2 +1} - 1) + \frac{\sqrt{x^2 + 1}}{2} {\rm arccoth} \sqrt{x^2 + 1} + \onehalf \sqrt{1 - x^2} \arctan {\rm h} \sqrt{1 - x^2} & (x < 1).
\end{array} \right. \nonumber
\end{equation}
The equivalent factor for the NFW profile,
$a_{{\rm NFW}}(x)$, is given by Bartelmann (1996). 

With $\beta \equiv D_L \psi / r_s$, $\alpha \equiv D_L \theta / r_s$, one can write the dimensionless lens equation:
\begin{equation}
\beta(\alpha) = \alpha - \lambda \frac{a(|\alpha|)}{\alpha}
\end{equation}
with
\begin{equation}
\lambda = 16 \pi \frac{G \rho_s r_s}{c^2} \frac{D_L D_{LS}}{D_S} = 7.21 h^{-1} \left( \frac{\rho_s}{{\rm M_{\odot} pc^{-3}}}\right) \left(\frac{r_s}{{\rm kpc}}\right) \frac{d_L d_{LS}}{d_S},
\end{equation}
where the reduced angular-diameter distance $d_A$ is defined to be $d_A
\equiv D_A H_0 / c$. In this letter, we use the angular diameter
distance of the Friedmann-Robertson-Walker universe.

\section{Lensing Properties and Implications}

The parameter $\lambda$ determines all the properties of the Burkert-profile lens. 
Fig.  1 shows the lens equations for different
values of $\lambda$. For $0 < \lambda \leq \lambda_c = 8 / \pi$, Burkert
halos cannot induce strong lensing. In contrast, the NFW halos can induce strong
lensing for any $\lambda$. Fig. 2 shows magnification $\mu$ of the
Burkert halos for different values of $\lambda$: $\mu = d \alpha^2 /
d\beta^2$. For $\lambda = 4 > \lambda_c$, two peaks in the
magnification curve correspond to the radial and tangential arcs
respectively. For $\lambda \leq \lambda_c$, the magnification curves
have the peak at $\alpha = 0$.

Modeling strong lensing with the Burkert halos requires $\lambda >
\lambda_c$, for which multiple images are formed only if 
the impact parameter $\beta \leq
\beta_c$. Here, $\beta_c$ is the maximum of $\beta(\alpha)$ in the
range $\alpha < 0$ (see Fig. 1). The cross section for strong lensing
by the Burkert halos is $\sigma = \pi (\beta_c r_s / D_L)^2$. If
$\beta = 0$, then an Einstein ring is formed at the radius $\alpha_t$,
which is a positive solution of $\alpha_t^2 = \lambda a(\alpha_t)$. 
(For a lens at $z_L = 0.5$ with a scale radius $r_s = 30 {\rm kpc}$, 
$\alpha_t = 0.1$ corresponds to $\Delta \theta \approx 1 \arcsec$.) 
For $ 0 < \beta < \beta_c$, three images are formed. The separation between the two
brighter images is $\Delta \theta \approx 2 \alpha_t r_s / D_L$. Fig. 3
shows the critical impact parameter $\beta_c$ and the Einstein radius
$\alpha_t$ as a function of $\lambda$ for 
the Burkert and NFW halos. 
While the impact parameter of the Burkert halo is always less than 
that of the NFW halo for fixed $\lambda$, the image separation 
$\alpha_t$ of the Burkert halos becomes larger than that of
the NFW halos when $\lambda$ becomes greater than $\approx 7$.

The difference between the two profiles near $r = 0$ leads to a dramatic
difference in the cross section for strong lensing for a given image
separation $\Delta \theta$ or $\alpha_t$. For small separations 
(say, $\alpha_t \lesssim 0.1$) the impact parameter $\beta_c$ required
for the Burkert halo to induce splitting is considerably smaller than that
of the NFW halo. For example, the ratios are $8.9 \times 10^{-4}$ and
$3.2 \times 10^{-2}$ for $\alpha_t = 0.01, 0.1$ respectively; the
relative cross sections scale as $\beta_c^2$. Thus for the same distribution
of $r_s$, the lensing optical depth due to Burkert halos is almost 
vanishingly small compared to that due to NFW halos. This is particularly
true for lensing by galaxy-scale objects; for example the probability for
a QSO image to be split by a single intervening giant elliptical galaxy
is at least $10^3$ times smaller for a Burkert profile than for an
NFW profile, for the same $r_s$ and mass. This is not true for large image separations ($\alpha_t \gtrsim 1$) as might be the case for
giant arcs formed by cluster-scale lenses. In this case the Burkert halos
yield optical depths comparable to the NFW halos, though slightly 
less (by a factor $\sim 0.1$ to $0.5$).

The condition $\lambda > \lambda_c$ corresponds to restricted area in
the ($\rho_s, r_s$) parameter plane. The condition can be rephrased
as:
\begin{equation}
 \left( \frac{\rho_s}{{\rm M_{\odot} pc^{-3}}}\right) \left(\frac{r_s}{{\rm kpc}}\right) > f(z_L, z_S)
\end{equation}
where 
\begin{equation}
f(z_L, z_S) = 0.353 h \frac{d_S}{d_L d_{LS}}.
\end{equation}
In Fig. 4 we plot strong lensing demarcation curves: $(\rho_s /{\rm
M_{\odot} pc^{-3}})(r_s / {\rm kpc}) = f_m(z_S)$ for $z_S = 1, 3, 5$.
Here, $f_m(z_S)$ is  the minimum of $f(z_L, z_S)$ in the range $0 < z_L
< z_S$. Burkert halos with ($\rho_s, r_s$) located above the
demarcation curve for a given $z_S$ can induce strong lensing, those
below the demarcation curve cannot.

The rotation curves of late-type LSB galaxies and dwarfs suggest that 
Burkert halos for these galaxies lie in a narrow strip in the ($\rho_s, r_s$) plane:
$\rho_s / {\rm M_{\odot} pc^{-3}} = 4.5 \times 10^{-2} (r_s / {\rm
kpc})^{-2 / 3}$, which is the thick solid line in Fig. 4. 
Salucci \& Burkert (2000) confirmed the relation for disk galaxies up to 100 times
more massive. At the highest masses in their sample there is an apparent change 
of slope, suggesting a cutoff in the masses
of halos that harbor individual disk galaxies at $M \sim 2 \times 10^{12} {\rm M_{\odot}}$.

We overplot some of these low-redshift measurements in Fig. 4. The black circles with
error bars are from the recent H$\alpha$ rotation curves of late-type
dwarf galaxies by Marchesini et al. (2002), which lie near the scaling
relation line. For comparison, we add the lensing cluster CL0024+1654 at
$z_L = 0.39$. The central density and scale radius of the cluster are 
estimated by
Firmani et al. (2001). Though the cluster is claimed to have soft core
(Tyson, Kochanski \& Dell'Antonio 1998; however see Broadhurst et al.
2000), there is no evidence that the cluster has the Burkert profile.
The cluster is clearly well off the scaling relation defined by low-redshift
disk galaxies (which, like the cluster, are argued to be dark-matter
dominated over most of their observed radii).

It is clear that the mass profiles of strong strong lens systems 
(e.g. Kochanek et al. 2003) must be cuspier than the Burkert
profiles on the spiral-galaxy scaling relation. If these systems
have Burkert profile DM halos, then the baryonic matter in these systems 
must be distributed in a much steeper way than the DM. It is then the
baryonic matter (not DM) that induces strong lensing. Unless the baryonic
matter has very steep core profiles, the total mass profile will remain
close to the Burkert profile and still be highly unlikely to induce
strong lensing. For example, the inclusion of exponential stellar disk
effectively reduces the $\lambda_c$ for face-on spirals:

\begin{equation}
\lambda_c = \frac{8}{\pi + 2 \Sigma_0 / \rho_s r_s},
\end{equation}
where $\Sigma_0$ is the surface density of the stellar disk at $\xi = 0$. 
For an $L^\star$ spiral with $M_{\rm disk} = 2 \times 10^{11} M_{\odot}$, we obtain $\Sigma_0 \approx 702 {\rm M_{\odot} pc^{-2}}$, the disk scale length $r_d \approx 4.76 {\rm kpc}$, $\rho_s \approx 3.78 \times 10^{-3} {\rm M_{\odot} pc^{-3}}$ and $r_s \approx 19.3 {\rm kpc}$ from the relation of Salucci \& Borriello (2002), who extend the scaling relation of Burkert halos including the exponential stellar disk. For this system (the rotational velocity $v_c = 240 {\rm km/s}$ at $r = 10 {\rm kpc}$), we have $\lambda_c \approx 0.357$, which is significantly less than $8 / \pi$. However, $\lambda$ is less than $0.18$ up to $z_S = 8$, which means that the face-on $L^\star$ spiral systems (Burkert halo + exponential stellar disk) cannot produce strong lensing up to $z_S = 8$, unless other baryonic matter such as a bulge is included.

Fig. 4 also shows that the Burkert halos extrapolated from the late-type
galaxy scaling relation cannot be used to account for lensing on
cluster scales ($r_s \gtrsim 100 {\rm kpc}$). 
One can model a lensing cluster with
a Burkert halo, whose $(\rho_s, r_s)$ should be beyond the demarcation
line for a given $z_S$ as those of the galaxy cluster CL0024+1654 are.
The cluster can induce strong lensing for $z_S > 1.0$, not for $z_S <
1.0$ if the cluster is indeed a Burkert halo. If DM halos at
cluster scales do have the Burkert profile, there must be be (at least)
two different families of the Burkert halos.

\section{Discussion}

We present an analytic lens equation for the Burkert halos, and use it to
characterize their lensing properties. We show that, as a result
of their soft core, the Burkert halos are much less
likely to induce strong lensing than the NFW halos.  While the qualitative
result is fairly obvious, when calculated in detail, the differences
in the strong-lensing predictions for the two profiles turn out to
be quite dramatic. If galaxies are indeed harbored by the Burkert halos, it implies that the baryonic matter (not DM) is responsible for strong lensing and that it is distributed in a much steeper way than the Burkert profile in strong lensing systems. For strong lensing systems (mostly elliptical galaxies), the total mass profiles for these systems have steeper
inner slopes than the Burkert profile (Koopmans \& Treu 2003). It remains to be seen whether
this is due to a distribution of baryons or cuspiness in the DM profile for these systems. 

On a galaxy scale, lens modeling combining stellar and DM components
has been carried out for the NFW profile (e.g.  Keeton 2001) and is
proving to be a useful tool to test the validity of the proposed DM
profile. Similar modeling with the Burkert profile will test its
validity. On a cluster scale, the validity of the Burkert halos can
be tested through galaxy kinematics (e.g. Sand, Treu \& Ellis 2002)
and X-ray observations (e.g. Wu \& Xue 2000). The detailed test methods
will be addressed in future articles. While challenging, it also
may be feasible to constrain halo profiles by combining measurements of
tangential shear and strong lensing for the same objects. 

\acknowledgments

We are grateful to Stefano Casertano for numerous conversations, sage
advice, and insightful comments on the manuscript.
Support for this work was provided by the Director's Discretionary
Research Fund at the Space Telescope Science Institute, which is operated
by AURA, Inc., under NASA contract NAS 5-26555.

\clearpage

\begin{figure}
\figurenum{1}
\plotone{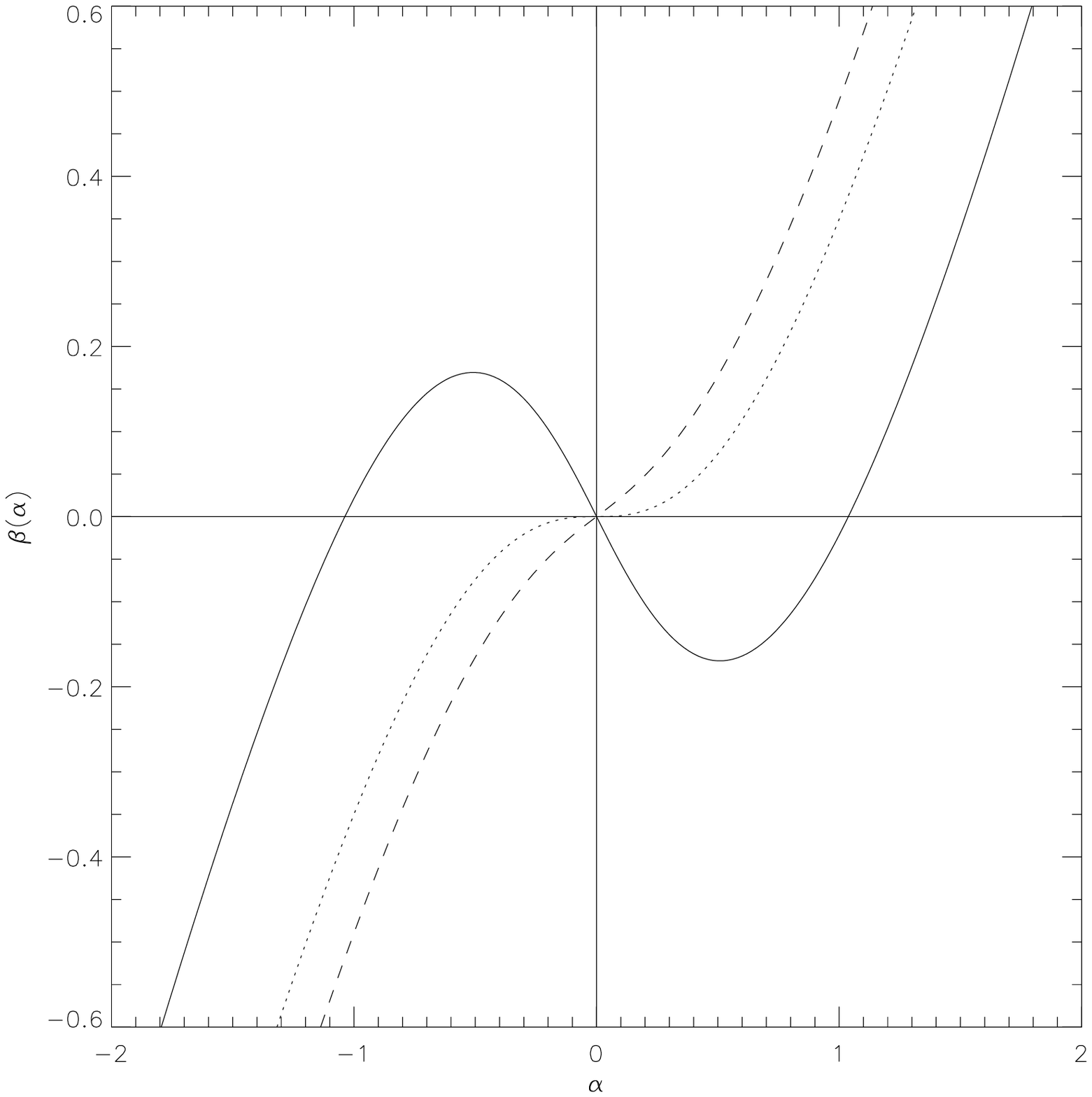}
\caption{Lens equation for the Burkert profile. 
The solid, dotted and dashed lines are respectively 
for $\lambda = 4, 8 / \pi, 2$. Only lens--source combinations
with $\lambda > 8 / \pi$ will produce multiple images. }
\end{figure}

\begin{figure}
\figurenum{2}
\plotone{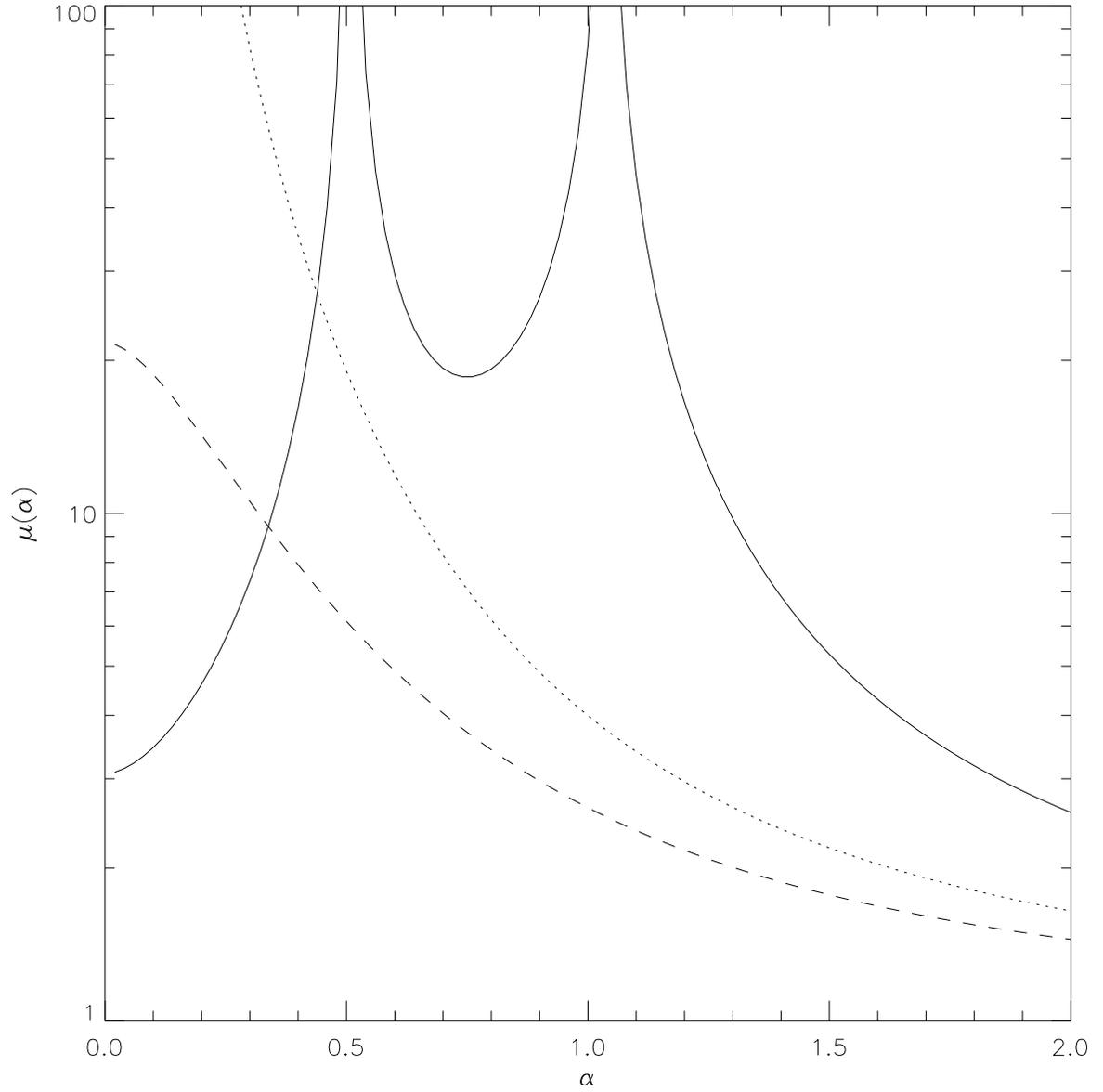}
\caption{Magnification as a function of image position 
for the Burkert profile. The solid, dotted and dashed lines are 
respectively for $\lambda = 4, 8 / \pi, 2$. The
inner and outer peaks in the magnification function for $\lambda = 4$
are for radial and tangential arcs, respectively.
}
\end{figure}

\begin{figure}
\figurenum{3}
\plotone{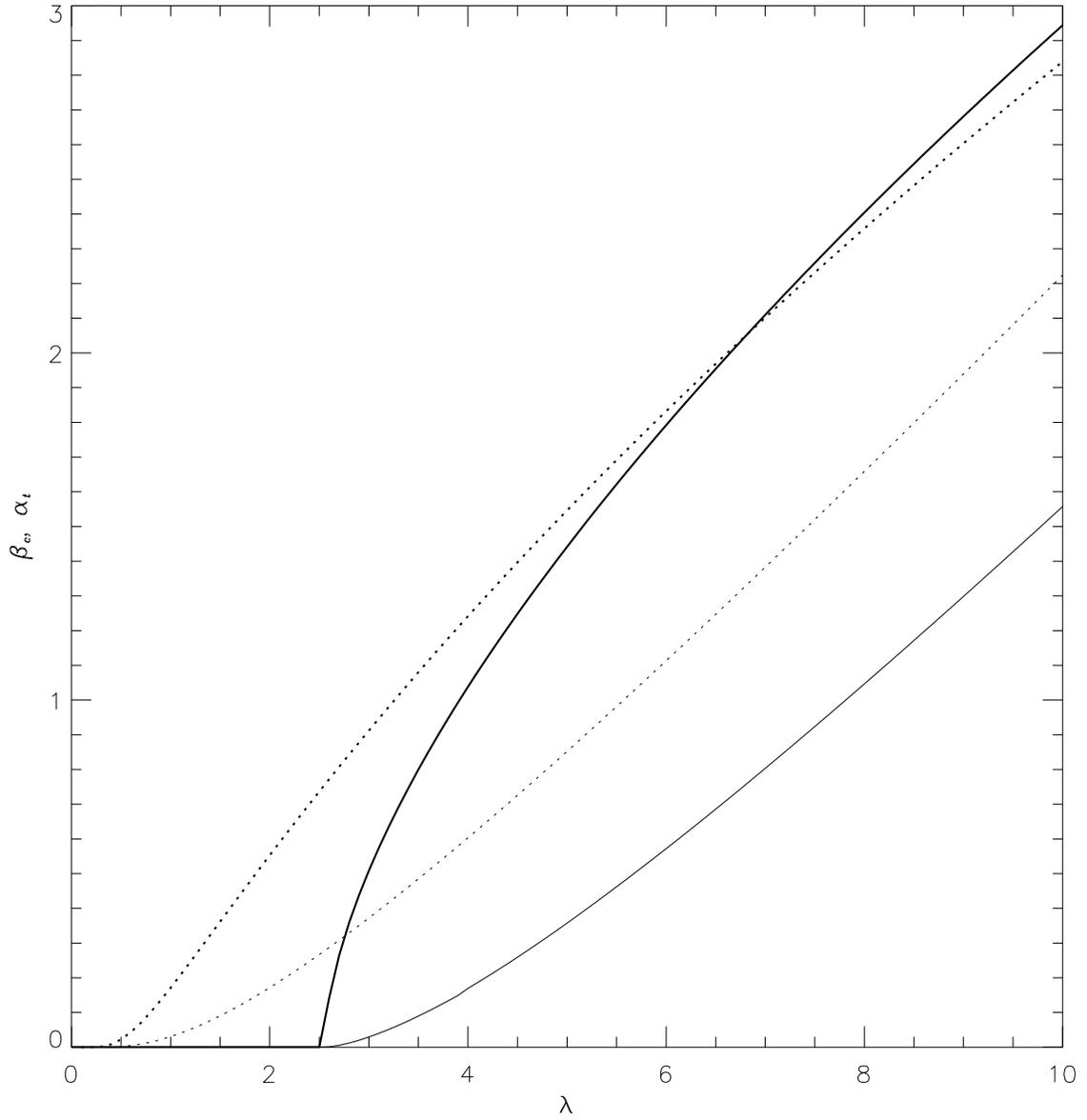}
\caption{The impact parameter $\beta_c$ (thin lines) and image
separation $\alpha_t$ (thick lines) as a function of $\lambda$ 
for the Burkert and NFW profiles. 
The solid and dotted curves are for the Burkert and NFW profiles, respectively.}
\end{figure}

\begin{figure}
\figurenum{4}
\plotone{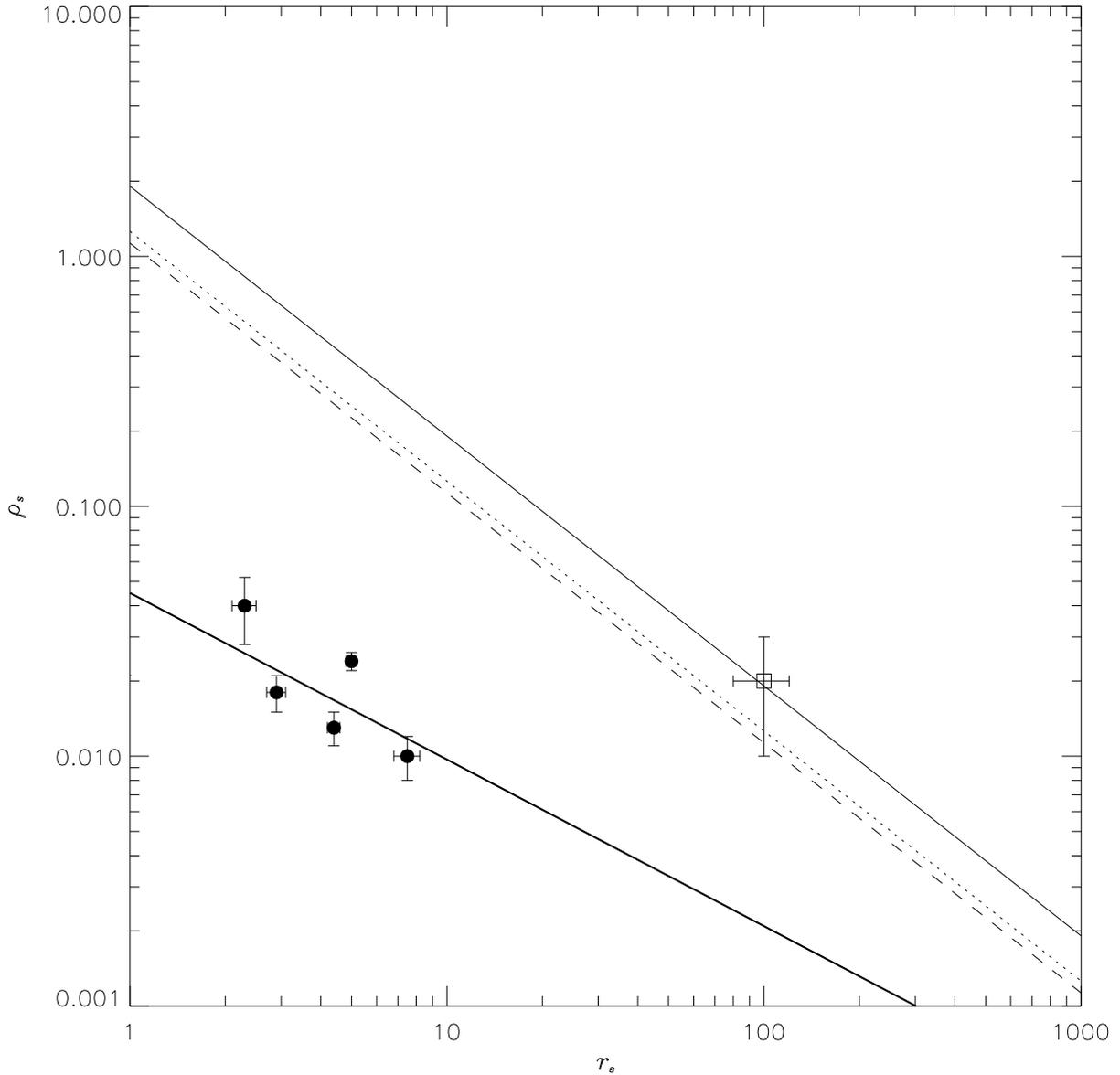}
\caption{Strong lensing demarcation curves for $z_S = 1, 3, 5$. 
The thin solid, dotted and dashed lines are respectively for source
redshifts $z_S = 1, 3, 5$. 
The thick solid line is the scaling relation of the Burkert halos 
at $z \approx 0$. 
The black circles with error bars are from rotation-curve measurements
of late-type low-surface-brightness galaxies and dwarfs (Marchesini et al. 2002). 
The square represents the halo parameters for the lensing
cluster CL0024+1654 estimated by Firmani et al. (2001). 
The units of $\rho_s$ and $r_s$ are ${\rm M_{\odot} pc^{-3}}$, kpc 
respectively.}
\end{figure} 

\end{document}